# The Clock and Control System for the ATLAS Liquid Argon Calorimeter Phase-I Upgrade

**Le Xiao,[a,b] Chonghan Liu,[b] Tiankuan Liu,[b,*] Hucheng Chen,[c] Jinghong Chen,[d] Kai Chen,[c] Yulang Feng,[d] Datao Gong,[b] Di Guo,[e,b] Huiqin He,[f,a,b] Suen Hou,[g] Guangming Huang,[a,*] Xiangming Sun,[a] Yuxuan Tang,[d] Ping-Kun Teng,[g] Annie C. Xiang,[b] Hao Xu,[c] Jingbo Ye,[b] Yang You[b]**

[a] *Central China Normal University,*
   *Wuhan, Hubei 430079, P.R. China*

[b] *Southern Methodist University,*
   *Dallas, TX 75275, USA*

[c] *Brookhaven National Laboratory,*
   *Upton, NY 11973, USA*

[d] *University of Houston, Houston,*
   *TX 77004, USA*

[e] *University of Science and Technology of China,*
   *Hefei Anhui 230026, China*

[f] *Shenzhen Polytechnic,*
   *Shenzhen 518055, China*

[g] *Academia Sinica,*
   *Nangang 11529, Taipei, Taiwan*
   *E-mail*: Guangming Huang (gmhuang@phy.ccnu.edu.cn) and Tiankuan Liu (tliu@mail.smu.edu)

ABSTRACT: A Liquid-argon Trigger Digitizer Board (LTDB) is being developed to upgrade the ATLAS Liquid Argon Calorimeter Phase-I trigger electronics. The LTDB located at the front end needs to obtain the clock signals and be configured and monitored remotely from the back end. A clock and control system is being developed for the LTDB and the major functions of the system have been evaluated. The design and evaluation of the clock and control system are presented in this paper.

# Contents

## 1. Introduction

A Liquid Argon Calorimeter (LAr) Trigger Digitizer Board (LTDB) is being developed to upgrade the ATLAS LAr trigger electronics, which will be commissioned in 2018-2019 [1]. Each LTDB includes 320-channel analog amplifiers, 80 Analog-to-Digital converter (ADC) ASICs [2], 20 serial-data transmitters (LOCx2's) [3], and 20 optical transmitter modules (MTx's) [4]. The ADC and LOCx2 each need a 40-MHz clock signal that is synchronized to the LHC bunch-crossing clock. The ADC, LOCx2, and MTx each need to be configured remotely after each power cycle. The operational status of the LTDB needs to be monitored. It is critical to design a clock and control system for the LTDB. In this paper, we present the design and the evaluation of the clock and control system for the LTDB.

     The remainder of the paper is organized as follows: Section 2 describes the overall design of the clock and control system for the LTDB and the design of the evaluation board. Sections 3-5 describe how each function of the clock and control system is implemented on the LTDB and how each function is evaluated. Section 6 summarizes the paper.

## 2. Overview of the design and evaluation

### 2.1 Design of the clock and control system

Figure 1 shows the block diagram of the clock and control system. The clock and control system uses five duplex optical links between each LTDB at the front end and the Front-End Link Interface eXchange (FELIX) board [5] at the back end. Each duplex link that serves 1/5-slice LTDB includes an optical transceiver module MTRx, a transceiver ASIC GBTX [6], and a GBT Slow Control Adapter (SCA) [7] on the LTDB. The GBTX recovers 40-MHz clocks from the received serial data and provides the recovered clocks to the ADCs and the LOCx2's. The GBTX also generates a Bunch-Crossing Reset (BCR) signal for each LOCx2. The GBT SCA configures all ASICs (ADCs, LOCx2, the dual-channel laser driver LOCld2 used in MTx, and the single-channel laser driver LOCld1 used in MTRx) after each power cycle and reads back the configuration and status at any moment. The GBT SCA controls the power supply modules and monitors the power supply status. To improve its reliability, each GBT SCA is connected to two GBTXs: one on the corresponding slice and the other on an adjacent slice. The optical links use commercial multi-channel optical transceivers: the MicroPOD produced by Avago Technologies and the Multi-Gigabit-Transceiver-embedded FPGAs on the FELIX.

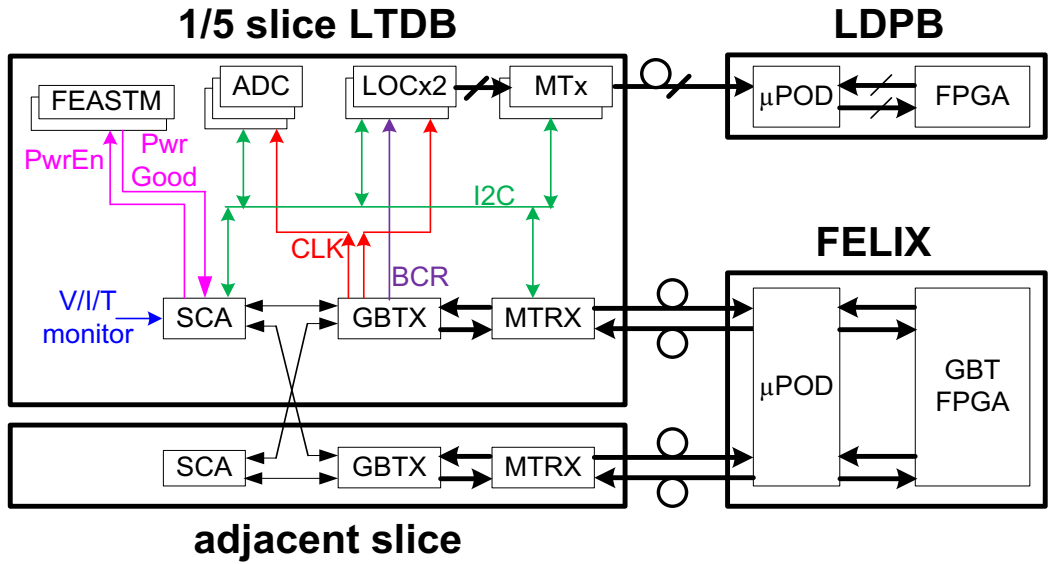

**Figure 1:** Overall design of the clock and control system.

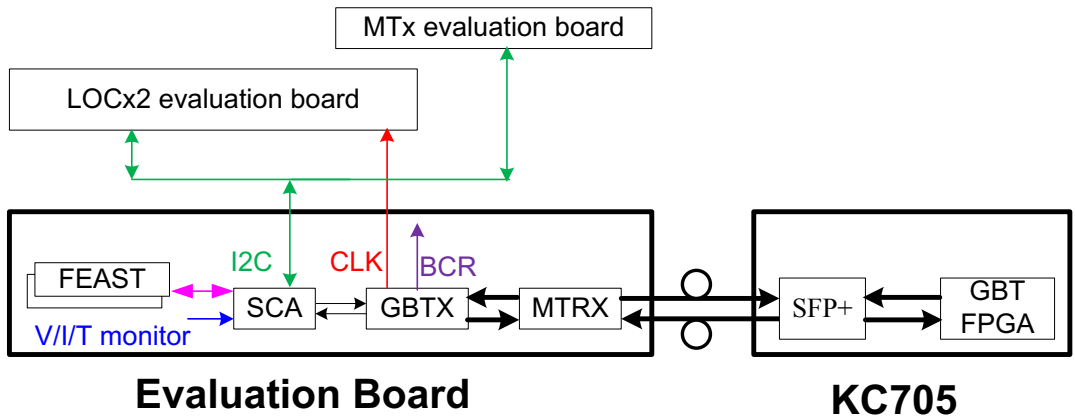

**Figure 2:** Block diagram of the evaluation system.

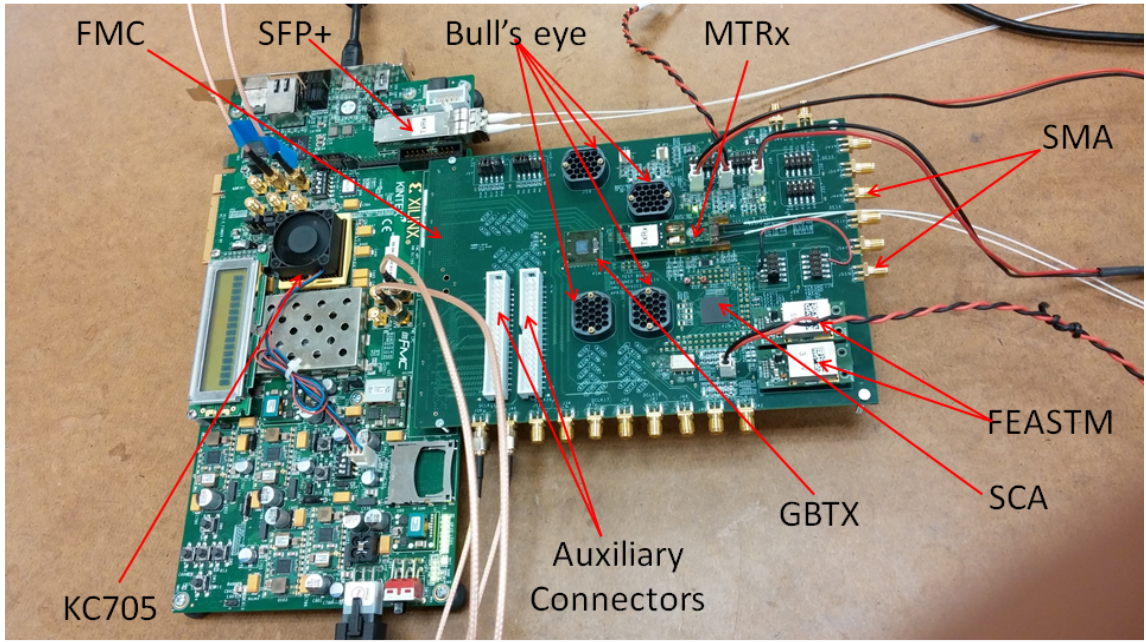

**Figure 3:** Picture of the evaluation board and KC705.

## 2.2 Design of the evaluation board

An evaluation board, including an MTRx, a GBTX, and a GBT SCA, has been developed to evaluate each function of the clock and control system. Figure 2 is the block diagram of the evaluation board. The evaluation board works together with a Xilinx Kintex FPGA board KC705 and also with standalone evaluation boards of LOCx2 and MTx. KC705 emulates the FELIX operating at the back-end. Figure 3 is a picture of the evaluation board and KC705.

## 3. Distribution of clocks and bunch-crossing resets

### 3.1 Design of the clock distribution

Each GBTX has 49 clock outputs that are grouped into three types. The first type includes 40 fixed-phase e-port clocks, labeled from dClk0 to dClk39. The second type includes eight programmable-phase clocks, labeled from ClockDES0 to ClockDES7. The third type is a dedicated e-port clock for the GBT SCA. Based on the evaluation that we will discuss in Section 3.2, we choose 20 fixed-phase clocks for 16 ADCs and 4 LOCx2's. The block diagram of the clock distribution is shown in red in figure 4.

### 3.2 Evaluation of clock distribution

The jitter of the recovered clocks of GBTX needs to be studied carefully since both the ADC and LOCx2 are sensitive to the clock jitter. A GBTX recovered clock is fed into a LOCx2 and the eye diagram of the serial output data is measured. The eye diagrams of LOCx2 with dClk17 and ClockDES0 as the reference clock are shown in figure 5. The eye diagram with dClk17 is much better than that with ClockDES0. The evaluation of using the GBTX recovered clock for the ADCs is still under study.

### 3.3 Distribution of bunch-crossing resets

Each LOCx2 implements a Bunch Crossing IDentification (BCID). All BCIDs need a common reset signal, the BCR, for the alignment of channels. The BCR signal repeats every 3564 LHC clock cycles and lasts for one LHC clock cycle. We use an e-port data output (dOut) of GBTX to provide the BCR signal to each LOCx2. In total, we need four out of 40 e-port data outputs of each GBTX. The block diagram of BRC generation is shown in purple in figure 4. We have observed the BCR signal generated in the evaluation board.

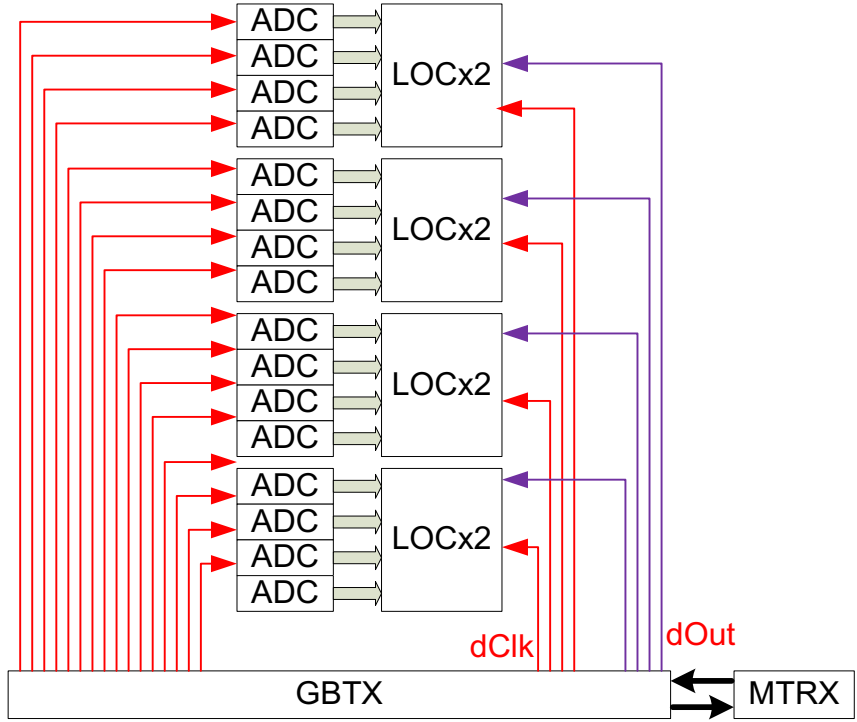

**Figure 4:** Block diagram of the distribution of clocks and BCRs.

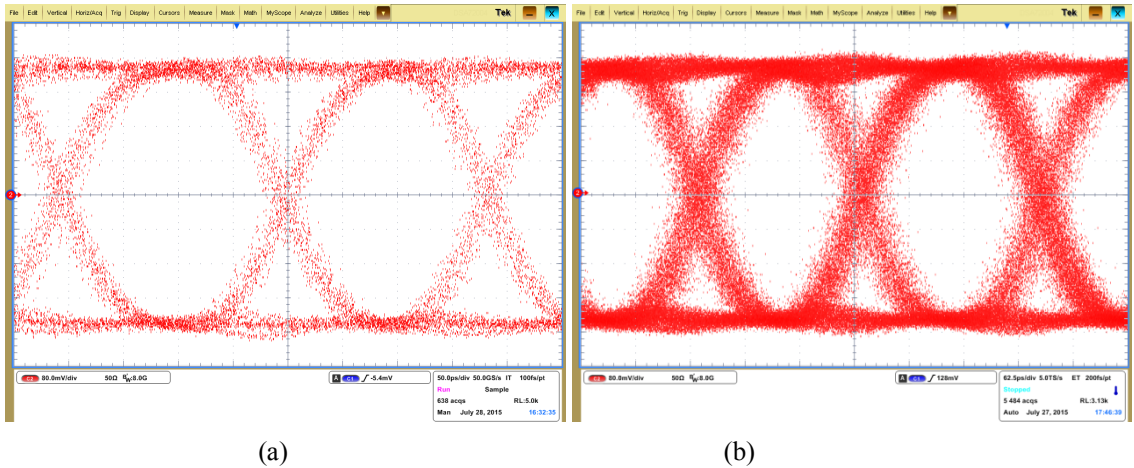

**Figure 5:** Eye diagrams of LOCx2 with (a) dClk17 and (b) ClockDES0 as the ref clock.

## 4. Configuration and status monitoring of ASICs

### 4.1 Design of the configuration and status monitoring of ASICs

We configure the internal registers of all ASICs, including the ADC, LOCx2, LOCld2 used in MTx, and LOCld1 used in MTRx, after each power cycle. The internal registers of the ASICs can be read back at any moment to monitor the operational status of these ASICs. We configure and monitor all ASICs via the I$^2$C masters of the GBT SCA. Each ASIC has an I$^2$C slave module, while each GBT SCA has 16 I$^2$C master modules. The block diagram of the ASIC configuration is shown in figure 6. Four ADCs connected to a LOCx2 share an I$^2$C master. The connected LOCx2 and the MTx share an I$^2$C master. The MTRx is configured via an extra I$^2$C master. In total, we use nine out of 16 I$^2$C masters of each GBT SCA.

### 4.2 Evaluation of configuration and status monitoring of ASICs

We have used the GBT SCA to configure successfully LOCx2, LOCld2, and LOCld1. Note that the GBT SCA is powered at 1.5 V, while the I$^2$C interfaces of LOCx2, LOCld2 and LOCld1 are all powered at 2.5 V. Furthermore, the ADC is powered at 1.2 V. All I$^2$C signals of the GBT SCA are pulled up to 1.5 V. It is guaranteed from the design that LOCx2, LOCld2, and LOCld1 all identify 1.5 V as active Logic High. The I$^2$C interface of the ADC can safely be pulled to 1.5 V. The interface between the GBT SCA and the ADC will be tested in the 1/5-slice LTDB in November 2015.

Note that the LOCld1 of the MTRx is not ready before the MTRx is configured. We have confirmed that the GBT SCA can configure MTRx when the uplink is not ready. It is convenient for the MTRx to have an extra I$^2$C connector and a switch between the GBT SCA I$^2$C and the external I$^2$C master so that we can configure MTRx before we set the phase aligner of GBTX for the GBT SCA.

## 5. Design and evaluation of the control and monitoring of power supplies

### 5.1 Control of power supply modules

The LTDB is capable of turning on or off all the power supply modules on the LTDB except the one used for the GBTX, the GBT SCA, and the MTRx. The LTDB uses two types of power supply modules. The analog components including amplifiers and ADCs use linear regulators such as LHC4913 [8] and LHC7913 [9]. The digital components including all LOCx2's and MTx's use DC-DC converters such as FEASTMP [10]. We use a general-purpose output of the GBT SCA to turn on or off FEASTMP, LHC4913, and LHC7913. The inhibit pin of LHC4913/LHC7913 is a TTL signal, whose Logic High level is beyond the voltage range (0 - 1.5 V) of the general-purpose output of the GBT SCA. Therefore, a bipolar transistor is used to translate the TTL signal to the desired voltage level. The transistor has been tested to meet the radiation tolerance requirements of ATLAS LAr Phase-I upgrade. We have confirmed in the evaluation that we can turn on or off the FEASTMP. The evaluation of controlling LHC4913/LHC7913 will be completed in November of 2015.

### 5.2 Monitoring of power supplies

We use four approaches to monitor the power supplies. First, we use a general-purpose input of the GBT SCA to observe the Power-Good signal of FEASTMP and the Over-Current-Monitor (OCM) signal of LHC4913 and LHC7913. The voltage range of the OCM signal of LHC4913 and LHC7913 is beyond the input voltage range of the general-purpose input pins of the GBT

SCA, so we use two resistors to divide the monitored voltages. We have confirmed in the evaluation that we can monitor the Power-Good signal of FEASTMP by using a general-purpose input of the GBT SCA. The control and monitoring of LHC4913/7913 will be evaluated in November of 2015.

Second, we use the ADCs of the GBT SCA to monitor the input voltages of FEASTMP, LHC4913, and LHC7913. The input voltage range of the ADC of the GBT SCA is from 0 to 1 V. For a power supply higher than 1 V, for example +5 V, we use two resistors to divide the monitored voltage into the ADC input range as shown in figure 7(a). For a negative voltage, we use two resistors and a positive voltage to generate a voltage within the range from 0 to 1 V. Figure 7(b) shows how we monitor the -5 V. Since the divided voltage shown in Figure 7(b) depends on both the positive and negative voltages, the approach shown in figure 7(b) is always combined with that shown in figure 7(a). We have confirmed on the evaluation board that we can use the ADCs of the GBT SCA to monitor the voltages. The voltages measured by using the ADC of the GBT SCA are consistent with those measured by using a digital multimeter.

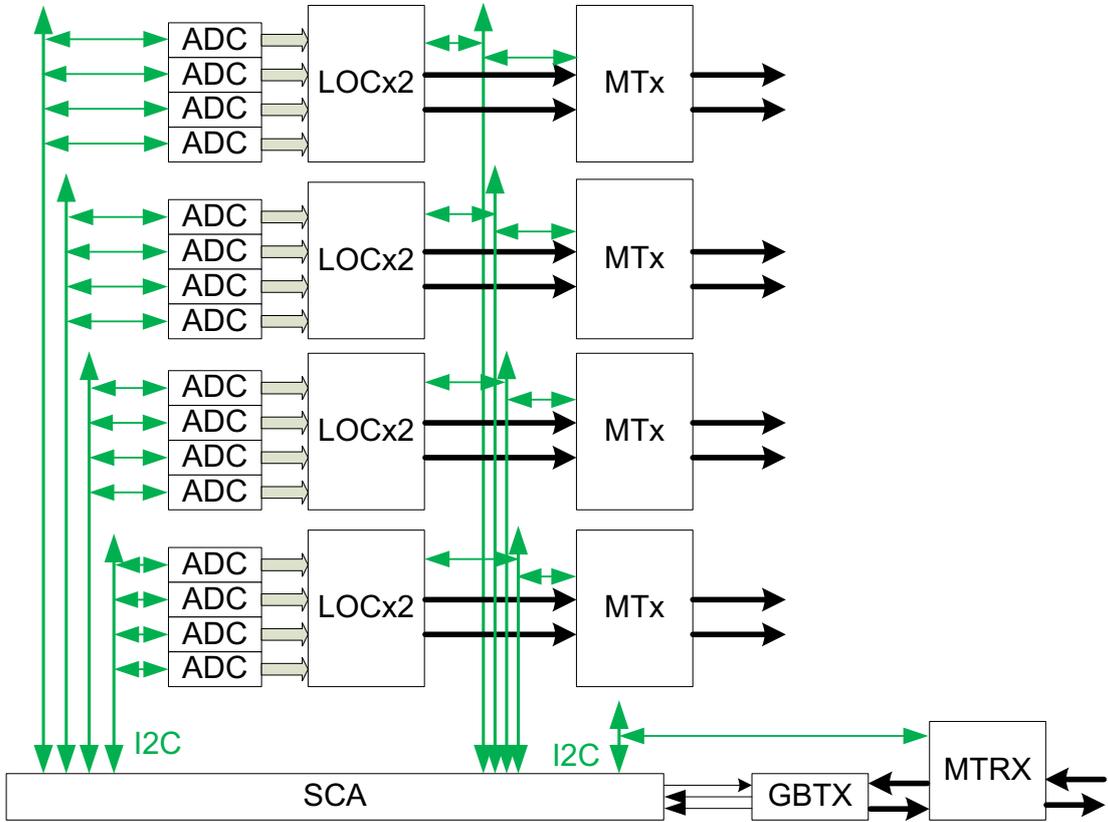

**Figure 6:** Block diagram of system configuration.

Third, we monitor the power supply currents by measuring the voltage drop on a sensing resistor. Since the GBT SCA cannot measure differential voltage directly, we have to use two ADC channels of the GBT SCA to measure a voltage drop. Figure 7(c) shows the schematic of the current monitoring of +5 V. In this example, R5 is a small sensing resistor on which the voltage drop is 0.1 V. R6, R7, R8, and R9 are dividing resistors to generate a voltage within the range from 0 to 1 V. Since the voltage drop on the sensing resistor is small, the dividing resistors (R6 - R9) must be precise (at least 0.1%) in order to achieve a reasonable precision (e.g. 10%). We have confirmed the current monitoring approach on the evaluation board. The currents measured by using the ADCs of the GBT SCA are consistent with those values measured by using a digital multimeter.

Finally, we use thermistors and the ADCs of the GBT SCA to monitor the temperatures at various location of LTDB. The GBT SCA has an embedded 100-µA current source. We have confirmed in the evaluation that we can use thermistors to monitor the temperatures.

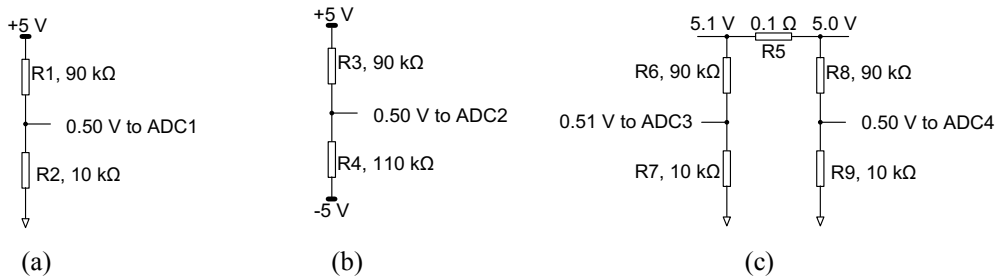

(a)  (b)  (c)

**Figure 7:** Schematics to monitor (a) a positive voltage higher than 1 V, (b) a negative voltage, and (c) a current.

## 6. Conclusion

A clock and control system is being developed to distribute the clock signals and control and monitor the LTDB remotely. A prototype of the clock and control system has been evaluated.

## Acknowledgments

This work is supported by the US-ATLAS R&D program for the upgrade of the LHC, the US Department of Energy under the Grant No. DE-FG02-04ER1299, and the National Natural Science Foundation of China under Grant No. 11375073 and 11420101004. The authors would like to express their deepest appreciation for Drs. Paulo Moreira, Pedro Leitao, Kostas Kloukinas, Alessandro Caratelli, Francois Vasey, and Jan Troska for kindly helping during the design and evaluation.